\documentclass[a4paper]{article}

\usepackage{INTERSPEECH2019}
\usepackage{multirow, makecell}
\usepackage{amssymb, amsmath, bm, mathtools,array}
\usepackage{color}
\newcommand{\bs}[1]{\boldsymbol{#1}}

\title{Initial investigation of an encoder-decoder end-to-end TTS framework using marginalization of monotonic hard latent alignments}
\name{Yusuke Yasuda$^{1,3}$, Xin Wang$^1$, Junichi Yamagishi$^{1,2,3}$}
\address{$^1$National Institute of Informatics, Japan
~~~$^2$The University of Edinburgh, Edinburgh, UK\\
$^3$SOKENDAI (The Graduate University for Advanced Studies), Japan}
\email{yasuda@nii.ac.jp, wangxin@nii.ac.jp,  jyamagis@nii.ac.jp}

\begin{document}

\maketitle
\begin{abstract}
End-to-end text-to-speech (TTS) synthesis is a method that directly converts input text to output acoustic features using a single network. A recent advance of end-to-end TTS is due to a key technique called attention mechanisms, and all successful methods proposed so far have been based on soft attention mechanisms. However, although network structures are becoming increasingly complex, end-to-end TTS systems with soft attention mechanisms may still fail to learn and to predict accurate alignment between the input and output. This may be because the soft attention mechanisms are too flexible.  Therefore, we propose an approach that has more explicit but natural constraints suitable for speech signals to make alignment learning and prediction of end-to-end TTS systems more robust. The proposed system, with the constrained alignment scheme borrowed from segment-to-segment neural transduction (SSNT), directly calculates the joint probability of acoustic features and alignment given an input text. The alignment is designed to be hard and monotonically increase by considering the speech nature, and it is treated as a latent variable and marginalized during training. During prediction, both the alignment and acoustic features can be generated from the probabilistic distributions. The advantages of our approach are that we can simplify many modules for the soft attention and that we can train the end-to-end TTS model using a single likelihood function.
%we can avoid unacceptable errors such as skipping input phones and repeating the same input segments. 
As far as we know, our approach is the first end-to-end TTS without a soft attention mechanism.
\end{abstract}

\section{Introduction}

End-to-end text-to-speech (TTS) synthesis method directly converts an input letter or phoneme sequence to an output acoustic feature sequence. All methods proposed so far have been based on an encoder-decoder sequence-to-sequence model with a soft attention mechanism \cite{Sotelo2017Char2wavES, Wang2017, DBLP:conf/iclr/TaigmanWPN18, DBLP:conf/iclr/PingPGAKNRM18, Shen2017, DBLP:journals/corr/abs-1809-08895}.
%Recent end-to-end TTS systems can produce synthetic speech with a quality comparable to that of  human speech \cite{Shen2017, DBLP:journals/corr/abs-1809-08895}. 
%All methods proposed so far are based on sequence-to-sequence model with soft attention mechanism \cite{Sotelo2017Char2wavES, Wang2017, DBLP:conf/iclr/TaigmanWPN18, DBLP:conf/iclr/PingPGAKNRM18, Shen2017, DBLP:journals/corr/abs-1809-08895}.

The aforementioned frameworks are very promising, and end-to-end TTS systems such as Tacotron 2~\cite{Shen2017} can produce synthetic speech with a quality comparable to that of human speech. The use of neural waveform models is one of the reasons, but we believe that improved model architectures are another important reasons for the improved performance of the end-to-end TTS systems. For example, a transition from Tacotron~\cite{Wang2017} to Tacotron 2~\cite{Shen2017} has extended the attention mechanism from additive attention~\cite{DBLP:journals/corr/BahdanauCB14} to location sensitive attention~\cite{DBLP:conf/nips/ChorowskiBSCB15}, post-net to improve predicted acoustic features, an additional stop flag prediction network, non-deterministic output generation using dropout during inference, and a significant increase in model parameter size. However, even if the attention network is well-trained, it may produce unacceptable errors such as skipping input words, repeating the same phrases, and prolonging the same sounds. 

Therefore, efforts have been made to improve or constrain soft attention mechanisms for end-to-end TTS systems. Most of them aim to enforce a monotonic alignment structure in order to reduce alignment errors. Such attempts include location-sensitive attention~\cite{DBLP:conf/nips/ChorowskiBSCB15}, monotonic attention~\cite{DBLP:journals/corr/Graves13}, and forward attention~\cite{Zhang2018}. In addition, techniques other than the attention mechanism itself can enforce monotonic alignment, such as window masking~\cite{DBLP:conf/iclr/PingPGAKNRM18, Zhang2018} and penalty loss for off-diagonal attention distribution \cite{DBLP:conf/icassp/TachibanaUA18}. Thus, although the quality of synthetic speech generated from end-to-end TTS systems is very high, the model architecture and its objective function are very complex, and many engineering tricks are used. 

Can we construct an encoder-decoder end-to-end TTS system without using such complicated networks and engineering tricks? We hypothesize that the main cause is the soft attention mechanisms. 
%For instance, context vectors of attention network are allowed to use input information at any time step of the encoder network. 
Therefore, we propose a new end-to-end TTS system that can be optimized based on a likelihood function only. Furthermore, the proposed TTS system uses more explicit but natural constraints for speech signals instead of soft attention mechanisms and should make alignment learning more robust and efficient. The constrained alignment is conceptually borrowed from segment-to-segment neural transduction (SSNT)~\cite{DBLP:conf/emnlp/YuBB16, DBLP:conf/iclr/YuBDGK17} and is extended to continuous outputs \footnote{SSNT is a similar model to RNN transducer \cite{DBLP:journals/corr/abs-1211-3711}, a well-known method in speech recognition. A notable difference is that SSNT separates transition probability from output probability, which is a preferred feature for TTS which is a task to predict continuous output.}. Because the SSNT calculates the joint probability of an acoustic feature sequence and an alignment given the input text, we can compute the likelihood function by marginalizing all possible alignments. The alignments are hard and monotonic increase by definition.

%The proposed framework and likelihood function are similar to those of a hidden Markov model (HMM) since all the possible alignments are marginalized using the forward-backward algorithm over a trellis consisting of input labels and output spectrum. 
The proposed framework and likelihood function are similar to those of a hidden Markov model (HMM) because all the possible alignments can be marginalized using a forward-backward algorithm over a trellis consisting of input labels and the output spectrum. 
However, unlike the conventional HMM or its mixture density network-based HMM~\cite{Tokuda+2016}, its transition and output probabilities are computed using encoder-decoder autoregressive neural networks like Tacotron. In other words, an input sequence is nonlinearly transformed into an encoded vector, and both the output and transition probabilities are computed in a non-linear autoregressive manner in the SSNT-based TTS.  

Our system is still under development and the quality of synthetic speech is not perfect yet. However, we present our first evaluation on the performance of the current system in this paper. We focus on our system's description and its evaluation for a standard reading speech corpus in this paper. Its application to a verbal performance corpus is investigated in \cite{KatoS2019rakugo}, and it shows effectiveness of our system in that corpus.

The rest of the paper is structured as follows. In Section 2, we overview end-to-end TTS systems with soft attention. In Section 3, we describe the new encoder-decoder TTS using marginalization of monotonic hard alignments and likelihood-based learning. In Section 4, we show experimental results using the system under development, and we summarize our findings in Section 5.    

%However, those recent techniques to improve the end-to-end TTS systems are mainly from an engineering perspective. With this direction, the attention mechanism and loss function to train the model become more and more complicated. In addition, none of the attention mechanisms proposed so far can provide both monotonic alignment assurance and generate high quality speech.

%In this research, we take a different approach from mainstream attention-based methods to reformulate end-to-end TTS from two standpoints: 1) monotonic alignment structure by design to avoid alignment errors, and 2) probabilistic modeling of alignment rather than engineering approaches to simplify and unify end-to-end TTS system.

\begin{table*}[t]
  \caption{Summary of major end-to-end TTS methods. All existing methods use soft attention mechanism for implicit alignment learning.}
  \label{tab:Summary of end-to-end TTS systems}
  \centering
  \begin{tabular}{llllll}
    \toprule
    {\bfseries System} & {\bfseries Network} & {\bfseries Alignment} & {\bfseries Decoder output} & {\bfseries Post-net output} & {\bfseries Waveform synthesis}\\
    \midrule
    Char2Wav \cite{Sotelo2017Char2wavES} & RNN & GMM & Vocoder & - & SampleRNN\\
    Tacotron \cite{Wang2017} & RNN & Additive & Mel & Linear & Griffin-Lim\\
    VoiceLoop \cite{DBLP:conf/iclr/TaigmanWPN18} & Memory buffer & GMM & Vocoder & - & WORLD\\
    Deep Voice 3 \cite{DBLP:conf/iclr/PingPGAKNRM18} & CNN & Dot-product & Mel & Linear/Vocoder & Griffin-Lim/WORLD/WaveNet\\
    Tacotron 2 \cite{Shen2017} & RNN & Location-sensitive & Mel & Mel & WaveNet\\
    Transformer \cite{DBLP:journals/corr/abs-1809-08895} & Self-attention & Dot-product & Mel & Mel & WaveNet\\\hline
    {\bfseries SSNT} (proposed) & {\bfseries RNN} & {\bfseries Hard} &  {\bfseries Mel} & {\bfseries -} & {\bfseries WaveNet} \\
    \bottomrule
  \end{tabular}
\end{table*}

\section{Overview of end-to-end TTS with soft attention}

Table~\ref{tab:Summary of end-to-end TTS systems} summarizes major end-to-end TTS methods. All the existing systems consists of an encoder and decoder with attention as basic components. Most systems have a pre-net~\cite{Wang2017} at the entrance of the decoder~\cite{Wang2017, DBLP:conf/iclr/PingPGAKNRM18, Shen2017, DBLP:journals/corr/abs-1809-08895}. Although some of the earliest studies~\cite{Sotelo2017Char2wavES, DBLP:conf/iclr/TaigmanWPN18} do not have the pre-net in the decoder, all the others do after \cite{Wang2017} introduced it.

The choice of target acoustic features is crucial for the end-to-end approach~\cite{yasuda-icassp2019}. Some of the earliest work chose vocoder parameters as a target~\cite{Sotelo2017Char2wavES, DBLP:conf/iclr/TaigmanWPN18}. Vocoder parameters are challenging because they have a long sequence length caused by fine analysis conditions for reliable feature extraction. \cite{Wang2017} used a mel spectrogram with coarse-grained analysis conditions as target features to reduce the gap of the length between input text and target acoustic feature sequences. All the other studies followed the same condition~\cite{Wang2017, DBLP:conf/iclr/PingPGAKNRM18, Shen2017, DBLP:journals/corr/abs-1809-08895}.

The attention mechanism is the core of those approaches. Many attention mechanisms have been used for TTS. Additive attention~\cite{DBLP:journals/corr/BahdanauCB14} and dot-product attention~\cite{ DBLP:conf/emnlp/LuongPM15} are content based families~\cite{DBLP:conf/nips/ChorowskiBSCB15} that consider input content to align input to output features. GMM attention~\cite{DBLP:journals/corr/Graves13} is a location based attention~\cite{DBLP:conf/nips/ChorowskiBSCB15} that consider input location only. Location sensitive attention~\cite{DBLP:conf/nips/ChorowskiBSCB15} extends the additive attention by considering the previous alignment, so it has both properties of content-based and location-based attention. The systems using vocoder parameters~\cite{Sotelo2017Char2wavES, DBLP:conf/iclr/TaigmanWPN18} seem to choose GMM attention. GMM attention has monotonic progress properties for modes of attention distributions for input location, so it is suitable for predicting long sequences like vocoder parameters. The systems based on CNN or self-attention to enable parallel training~\cite{DBLP:conf/iclr/PingPGAKNRM18, DBLP:journals/corr/abs-1809-08895} seem to use do-product attention which can be combined with positional encoding~\cite{Vaswani2017} for constructing sequential relations and an initial monotonic alignment in parallel. The systems based on RNN seem to use additive attention and its extension~\cite{Wang2017, Shen2017, Zhang2018, yasuda-icassp2019}.

In addition to the decoder, some systems have a post-net, an additional network that predicts acoustic features. A post-net was originally introduced to convert acoustic features to different acoustic features that were suitable for an adopted waveform synthesis method, for example, from mel spectrograms to linear spectrograms~\cite{Wang2017} or mel spectrograms to vocoder parameters~\cite{DBLP:conf/iclr/PingPGAKNRM18}. In recent studies the role of the post-net was to improve the acoustic features predicted by the decoder to improve quality further~\cite{Shen2017, DBLP:journals/corr/abs-1809-08895}. The post-net introduces an additional loss term in the objective function.

Recent methods predicts binary stop flags to determine when to stop prediction~\cite{Shen2017, DBLP:conf/iclr/PingPGAKNRM18, DBLP:journals/corr/abs-1809-08895}. As opposed to predicting the fixed length output, the stop flag enables avoiding unnecessary computation. The stop flag introduces an additional loss term in the objective function.

Speech is nondeterministic. An exactly identical utterance can not be reproduced in speech. To enable the randomness, \cite{Shen2017} enables dropout in pre-net during inference.

Although the aforementioned techniques have contributed to the advancement of the end-to-end TTS, some issues remain unsolved by these techniques. All content-based attentions and their extensions do not guarantee forward progress of alignment from the previous position. They include forward attention~\cite{Zhang2018}, which extends the location sensitive attention by incorporating monotonic transition formulation. The GMM attention~\cite{DBLP:journals/corr/Graves13} has monotonic progress properties for the mode position of each Gaussian component. However, it sometimes gives broad, mode-free, or multi-modal distributions that result in muffled speech, possibly because it does not consider the content of input. Monotonic attention~\cite{DBLP:conf/icml/RaffelLLWE17} is the only attention mechanism that guarantees monotonic progress by switching to hard attention during inference from soft attention during training, but it is not successful for text-to-speech synthesis~\cite{DBLP:conf/iclr/PingPGAKNRM18}.

Learning stop flags seems to be trivial because it is just a binary classification task and because only one boundary has a flag turn from not to stopping to stopping. However, it tends to overfit because its loss has an order of magnitude lower than the acoustic feature loss term. To alleviate this problem, \cite{DBLP:journals/corr/abs-1809-08895} introduced higher weighting at the boundary where a flag reaches a stop state in the binary loss term. The stop flag is an extra complexity to implement such a trivial feature.

Enabling randomness for generated speech by dropout is not widely utilized because the dropout is normally disabled during prediction.

\section{Proposed SSNT-based TTS}

\begin{figure}[t]
  \centering
  \includegraphics[width=\linewidth]{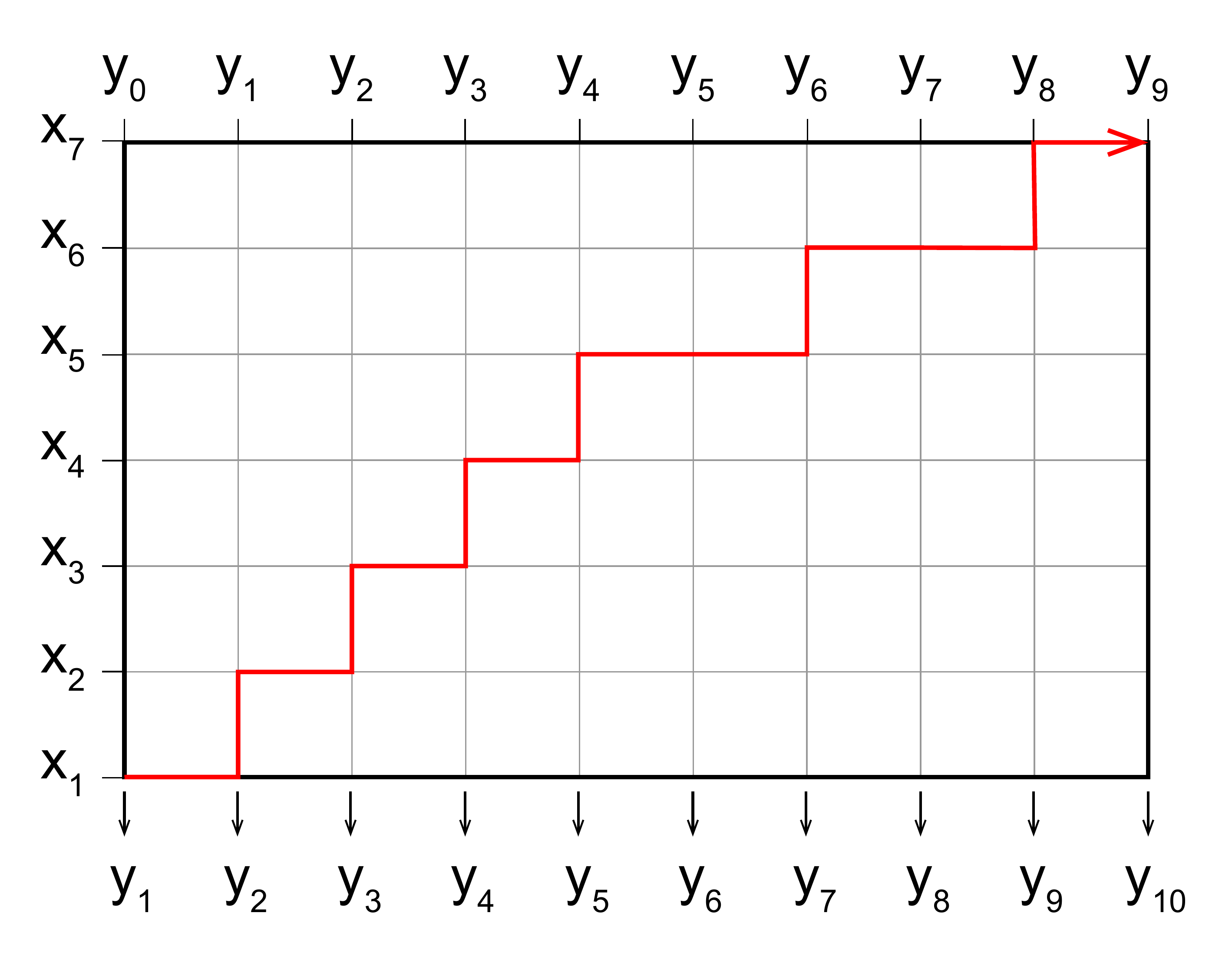}
  \caption{Trellis structure of our model. A path that connects $\bs{x}$ and $\bs{y}$ represents alignment.}
  \label{fig:ssnt-hard-alignment}
\end{figure}

\subsection{Model definition and learning of SSNT-based TTS}
Let us denote the input letter or phone sequence as $\bs{x}_{1:I}=\{x_{1}, \cdots  x_I\}$, where 
$x_i$ is a $i$-th letter or phone of the input sequence. We then use $\bs{y}_{1:J}=\{\bs{y}_1, \cdots, \bs{y}_J\}$ and $\bs{y}_j \in \mathbb{R}^{D}$ to denote an output acoustic feature sequence and acoustic features at time $j$, respectively. Our approach is to model the output acoustic feature sequence $\bs{y}_{1:J}$ given a letter or phone sequence $\bs{x}_{1:I}$ by marginalizing all possible alignments over a trellis consisting of the input and output sequences:
\begin{equation}
\label{ssnt_eq1}
p(\bs{y}_{1:J} \mid \bs{x}_{1:I}) = \sum_{\forall \mathbf{z}} p(\bs{y}_{1:J}, \mathbf{z} \mid \bs{x}_{1:I}),
\end{equation}
where $\mathbf{z} = \{z_{1}, \cdots  z_J\} = \{1, 1, , \cdots  I-1, I\}$ represents one of the possible paths of the trellis. Figure \ref{fig:ssnt-hard-alignment} shows trellis structure of our model.
%
%There are two possible marginalization methods that are tractable to compute the integral: 1) variational approximation \cite{DBLP:conf/icml/MnihG14, DBLP:conf/icml/MnihR16}, and 2) polynomial marginalization \cite{DBLP:conf/emnlp/YuBB16, DBLP:conf/emnlp/WuSC18}. The former method optimizes a lower bound on the log likelihood using samples from variational posterior. It allows to model flexible latent alignment structure, but the approximated lower bound may not be tight enough due to high variance of sampling, which may result in poorly trained model and low likelihoods. The latter method assumes restriction to its alignment structure, but it can exactly marginalize latent alignments and directly optimize a total likelihood. We take the latter approach with monotonic alignment structure with self transition, because text and speech data used for training TTS systems normaly have time-wise monotonic relationship.
%
%\subsection{Segment-to-segment neural transduction}
%
We then use the concept of SSNT~\cite{DBLP:conf/emnlp/YuBB16, DBLP:conf/iclr/YuBDGK17}. More specifically, we factorize the joint probability of Eq.~(\ref{ssnt_eq1}) into an alignment transition probability and emission probability for acoustic features with the 1st-order Markov assumption of $\mathbf{z}$:
\begin{align}
&p(\bs{y}_{1:J}, \mathbf{z} \mid \bs{x}_{1:I}) \approx \nonumber\\
&\prod_{j=1}^{J} p(z_j \mid z_{j-1}, \bs{y}_{1:j-1}, \bs{x}_{1:I}) p(\bs{y}_j \mid \bs{y}_{1:j-1}, z_j,  \bs{x}_{1:I}) \label{eqn:factorized-joint-prob}
\end{align}
and we use neural networks to compute the two probabilities. 
%In this paper, we use an encoder-decoder network with autoregressive feedback similar to Tacotron to compute the probabilities. 

%Note that the formulation assumes that the acoustic feature and alignment are conditioned on prefix of input linguistic feature. However we can relax the condition by using bidirectional RNN at encoder to consider future information of linguistic feature. 

To constrain the alignment probability as left-to-right with a self transition,  an alignment transition variable with two possible values, $a_{i,j} = \left\{\mathtt{Emit}, \mathtt{Shift}\right\}$ is further introduced for the alignment probability of Eq.~(\ref{eqn:factorized-joint-prob}): 
\begin{align}
\label{ssnt_tts_eq_transition_probability}
& p(z_j = i \mid z_{j-1}, \bs{y}_{1:j-1}, \bs{x}_{1:I}) = \nonumber\\
& \begin{cases}
0 & z_{j-1} > i \\
p(a_{i,j} = \mathtt{Emit}) & z_{j-1} = i \\
p(a_{i-1,j} = \mathtt{Shift})p(a_{i,j} = %\mathtt{Emit}) & z_{j-1} \neq i\\
\mathtt{Emit}) & z_{j-1} = i - 1\\
0 & z_{j-1}  < i - 1
\end{cases}
\end{align}
Note that the $\mathtt{Emit}$ transition keeps the input position, while $\mathtt{Shift}$ transition proceeds and reads one more input. Please also note that $p(a_{i,j} = \mathtt{Emit}) = 1 - p(a_{i,j} = \mathtt{Shift})$ and that a neural network predicts only one of them. 

%Using the probability of $a_{i,j}$, the alignment probability is computed as follows:

%\begin{equation}
%\label{ssnt_eq_transition_probability}
%    \begin{split}
%        p(z_j = i| z_{j-1}, \mathbf{y}_1^{j-1}, \mathbf{x}_i^i) = \\ %\begin{cases}
%0 & i < z_{j-1}\\
%p(a_{i,j} = \mathtt{Emit}) & i = z_{j-1}\\
%\left(\prod_{i'=z_{j-1}}^{i-1} p(a_{i',j} = \mathtt{Shift})\right)p(a_{i,j} %= \mathtt{Emit}) & i > z_{j-1}
%        \end{cases}
%    \end{split}
%\end{equation}

The emission probability was a discrete distribution because SSNT was originally proposed for NLP tasks such as abstractive sentence summarization, morphological inflection generation, and machine translation and because its outputs are words, the emission probability was a discrete distribution. In our case, the output is continuous, and hence we have to define our own emission probability distribution function. In this paper, we simply use a multivariate Gaussian distribution as the emission probability of the acoustic features: 
\begin{equation}
p(\bs{y}_j \mid \bs{y}_{1:j-1}, z_j, \bs{x}_{1:I}) =\mathcal{N}(\bs{y}_j ; \boldsymbol{\mu}, \sigma^2\mathbf{I})
\label{eqn:emission}
\end{equation} 
Note that $\boldsymbol{\mu}$ is predicted by an encoder-decoder network with autoregressive feedback similar to Tacotron. 

Our model can be trained by minimizing the negative log likelihood:
\begin{align}
\mathcal{L}(\boldsymbol{\theta}) &= -\log p(\bs{y}_{1:J} \mid \bs{x}_{1:I} ; \boldsymbol{\theta}) \nonumber\\
&= -\log\alpha(I, J)
\end{align}
Here $\alpha(I, J)$ is a forward variable of the forward-backward algorithm at the final input position $I$ and the final time step $J$. The final forward variable can be calculated recursively:
\begin{align}
    \text{For } j &= 1, \nonumber\\
    & \alpha(i, 1) = p(z_1 = 1 \mid \bs{x}_{1:I})p(y_1 \mid z_1, \bs{x}_{1:I}) \nonumber\\
    \text{For } j &> 1, \nonumber\\
    & \alpha(i, j) = p(y_j \mid \bs{y}_{1:j-1}, z_j,  \bs{x}_{1:I}) \cdot \nonumber\\
    & \bigl\{\alpha(i - 1, j - 1)p(z_j = i \mid z_{j-1} = i - 1, \bs{y}_{1:j-1}, \bs{x}_{1:I}) \nonumber\\
    & + \alpha(i, j - 1)p(z_j = i \mid z_{j-1} = i, \bs{y}_{1:j-1}, \bs{x}_{1:I})\bigr\}
\end{align}

The gradients of the negative log likelihood with respect to $\boldsymbol{\theta}$ can be computed using the standard back-propagation algorithm. For more efficient gradient computation, the gradient can be computed with the following form:

\begin{align}
&\frac{\partial \log p(\bs{y}_{1:J} \mid \bs{x}_{1:I} ; \boldsymbol{\theta})}{\partial\boldsymbol{\theta}} = \nonumber\\ &\frac{1}{p(\bs{y}_{1:J} \mid \bs{x}_{1:I} ; \boldsymbol{\theta})}\sum_{i=1}^{I}\sum_{j=1}^{J}\frac{\partial p(\bs{y}_{1:J} \mid \bs{x}_{1:I} ; \boldsymbol{\theta})}{\partial\alpha(i,j)}\frac{\partial\alpha(i,j)}{\partial\boldsymbol{\theta}}
\end{align}

Then the back-propagation can be combined via backward operation of the forward-backward algorithms by introducing a backward variable $\beta(i, j)$ and utilizing the relation $\displaystyle \frac{\partial p(\bs{y}_{1:J} \mid \bs{x}_{1:I} ; \boldsymbol{\theta})}{\partial \alpha(i, j)} = \beta(i, j)$ \cite{DBLP:conf/emnlp/YuBB16, DBLP:conf/emnlp/Eisner16}. 

\subsection{Inference of SSNT-based TTS}
\label{subsec:Text-to-speech synthesis based on SSNT}

During prediction, the alignment variable $z_j$ is incremented by sampling from the Bernoulli distribution with a parameter obtained by normalizing the two nonzero cases of Eq.~(\ref{ssnt_tts_eq_transition_probability}), namely, $z_{j} = z_{j-1} + k$, where $k \sim \mathrm{Bernoulli}(p)$ and 
\begin{equation*}
p = \frac{p(a_{i-1,j} = \mathtt{Shift})p(a_{i,j} = \mathtt{Emit})}{ p(a_{i,j} = \mathtt{Emit}) + p(a_{i-1,j} = \mathtt{Shift})p(a_{i,j} = \mathtt{Emit})}.
\end{equation*}

Note that this is different from a transition matrix in HMM synthesis which models duration with exponential distribution. Our system models transition of alignment as a latent variable so our system do not define distribution of duration like HMM synthesis.

In the original SSNT, its prediction stops when the EOS token is the output. In our case, its prediction stops when the alignment variable reaches the final input position\footnote{\cite{DBLP:conf/iclr/YuBDGK17} uses the same criteria to stop prediction to consume full input.} causing all of the input sequence to be used.

Acoustic features may be sampled from the Gaussian distribution using the conditional input at the sampled alignment position. However, in this paper, we simply use the mean of the prediction probability distribution as the generated acoustic features. We will investigate the random sampling strategy after we properly estimate a full covariance of Eq.~(\ref{eqn:emission}). 

\subsection{Network structure of SSNT-based TTS}

\begin{figure}[t]
  \centering
  \includegraphics[width=\linewidth]{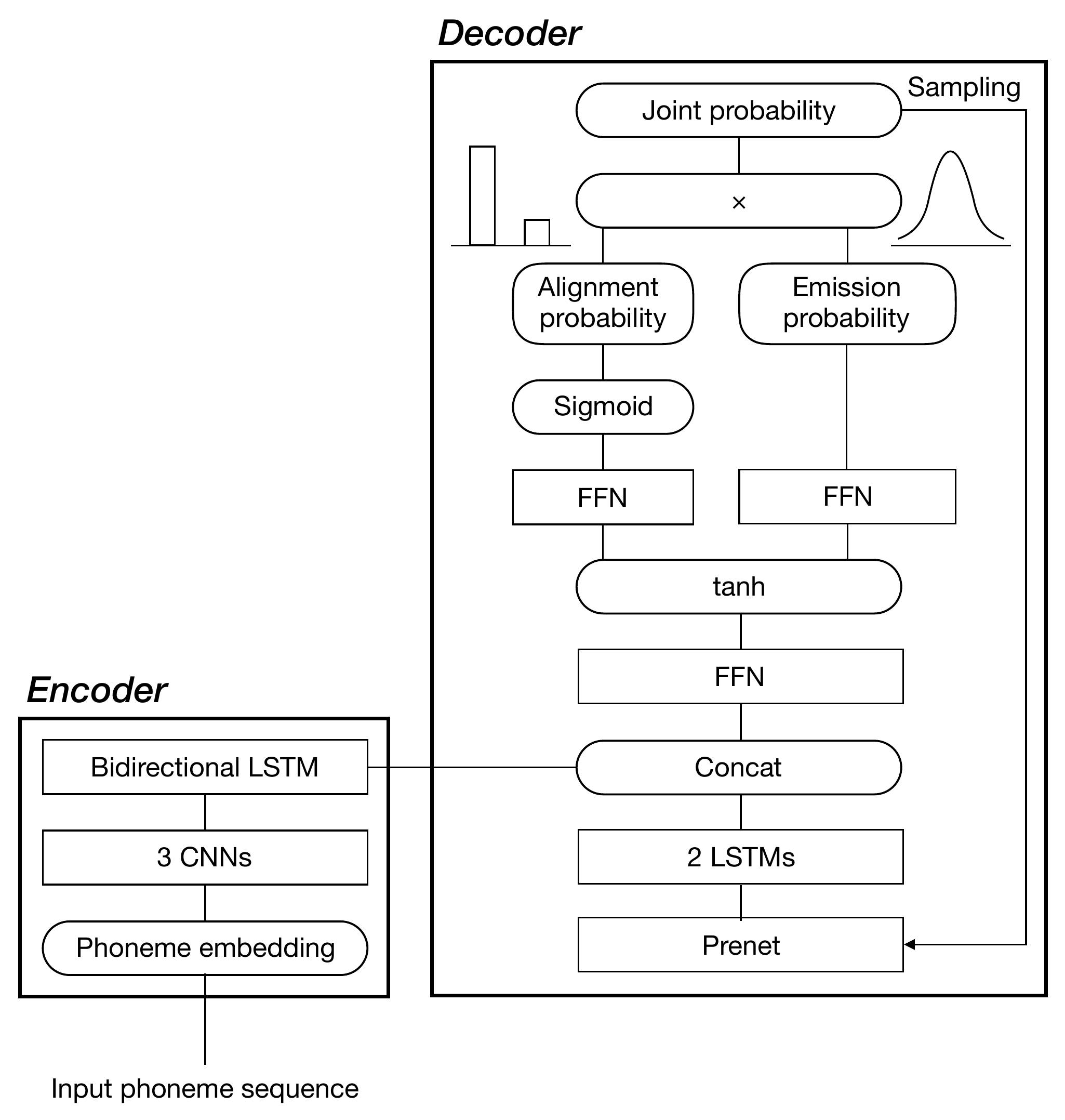}
  \caption{Detailed network structure of the SSNT-based TTS system.}
  \label{fig:ssnt-structure-old}
\end{figure}

Figure~\ref{fig:ssnt-structure-old} shows the detailed network structure of the system for SSNT-based speech synthesis system. The network consists of an encoder and decoder. The encoder processes either a letter or phone sequence. We then use the CNN-based encoder~\cite{Shen2017}, which consists of a stack of convolutional layers, and a bidirectional LSTM layer~\cite{DBLP:journals/tsp/SchusterP97}. 

The decoder processes the acoustic feature sequence in an autoregressive manner. The predicted acoustic features from the previous time step are first processed by the pre-net~\cite{Wang2017}, which consists of fully connected layers with ReLU activation~\cite{DBLP:conf/icml/NairH10} and dropout regularization~\cite{DBLP:journals/jmlr/SrivastavaHKSS14}. The pre-net's output is passed through a stack of LSTM layers~\cite{DBLP:journals/neco/HochreiterS97}. The LSTM stack's output is concatenated with the encoder's output and then transformed by a fully connected layer with tanh activation.  The output of tanh nonlinearity is passed to two networks. One is a fully connected layer with sigmoid activation to compute the alignment transition probability $p(a_{i,j})$ of Eq.~(\ref{ssnt_tts_eq_transition_probability}). The other is a linear layer to compute the emission probability of Eq.~(\ref{eqn:emission}). 

\section{Experiments}

\subsection{Experimental conditions}

We used the same conditions as in our previous experiment~\cite{yasuda-icassp2019}. A Japanese speech corpus from the ATR Ximera dataset~\cite{kawai2006ximera} was used. This corpus contains a total of 28,959 utterances from a professional female speaker and is around 46.9 hours in duration. We used manually annotated phonemes labels~\cite{Luong2018}. The phoneme labels had 58 classes, including silences, pauses, and short pauses. All sentences start and end with a silence symbol. Although Japanese is a pitch-accented language, we did not use accentual type labels in this paper. To train our proposed systems, we trimmed the beginning and ending silence from the utterances, after which the duration of the corpus was 33.5 hours. Fixed length silences were prepended and appended to target mel spectrogram. These data preparation made any phoneme symbols inappropriate to be skipped. The frame size and shift used for the mel spectrogram were 50\,ms and 12.5\,ms, respectively.  We used 27,999 utterances for training, 480 for validation, and 480 for testing.

Phoneme embedding vectors have 256 dimensions. For the encoder, we used the same conditions as \cite{Shen2017}. For the decoder, we used two pre-net layers with 256 and 128 dimensions, two LSTM layers with 256 dimensions each, and two fully connected layers for context projection with 256 dimensions each. We applied zoneout regularization~\cite{Krueger2016} to all LSTM layers with probability 0.1 as in \cite{Shen2017}. We set the reduction factor~\cite{Wang2017} to be 2. The Adam optimizer was used for training~\cite{DBLP:journals/corr/KingmaB14}. The validation loss was steady so we stopped training at 510 epochs by checking quality of predicted spectrogram in validation set. Finally, we used $\mu$-law WaveNet for the waveform generation~\cite{wavenet}.

\subsection{Subjective evaluations}

We conducted a listening test to measure the quality of synthetic speech. We chose Japanese Tacotron and self-attention Tacotron without accentual type labels~\cite{yasuda-icassp2019} as baselines in this experiment. All the synthetic speech waveforms were generated using the identical WaveNet model, which was trained using natural mel spectrograms with a 12.5\,ms frame shift and 16\,kHz sampling frequency.

We recruited 104 native speakers of Japanese as listeners by crowdsourcing. The listeners evaluated 40 audio samples that contained eight randomly selected sentences generated by each of five systems in a random order in a single test set. The systems included natural speech and analysis-by-synthesis in addition to our system and the two aforementioned baseline systems. One listener could evaluated at most ten test sets. One audio sample was evaluated five times, and we got a total of 19,200 data points.

\begin{figure}[t]
  \centering
  \includegraphics[width=\linewidth]{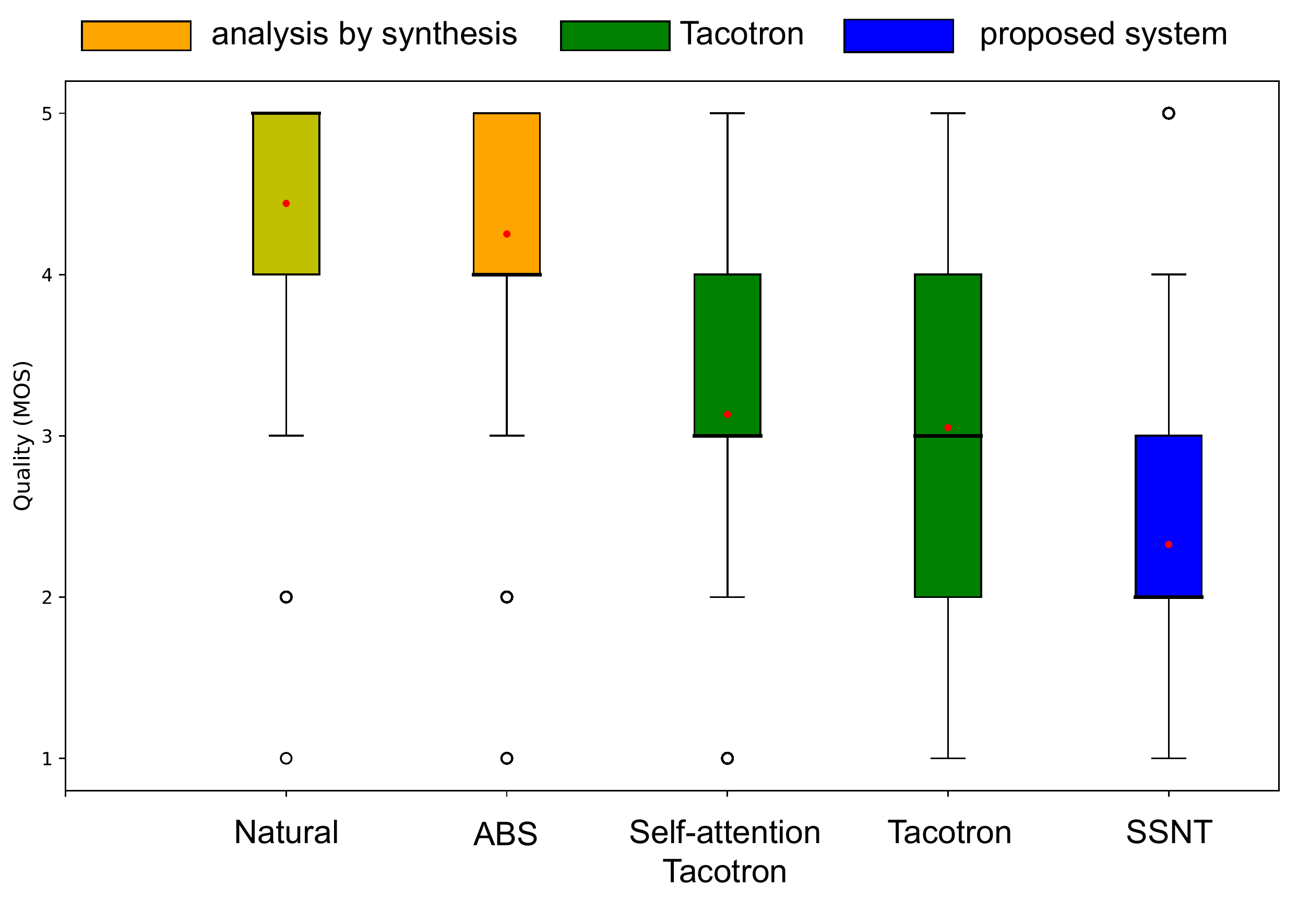}
  \caption{MOS scores of subjective evaluation.}
  \label{fig:listening_test}
\end{figure}

Figure~\ref{fig:listening_test} shows the results of the subjective evaluation. Among the baseline systems, self-attention Tacotron got 3.13$\pm$0.03, and Japanese Tacotron got 3.05$\pm$0.03. The scores of the baseline systems are consistent with the previous work~\cite{yasuda-icassp2019}. The scores were relatively low because we did not provide any pitch accentual type labels, even though Japanese is a pitch-accented language. Our system got a MOS score of 2.33$\pm$0.03. Unfortunately, it was rated lower than the baseline systems. 

\begin{figure}[t]
  \centering
  \includegraphics[width=\linewidth]{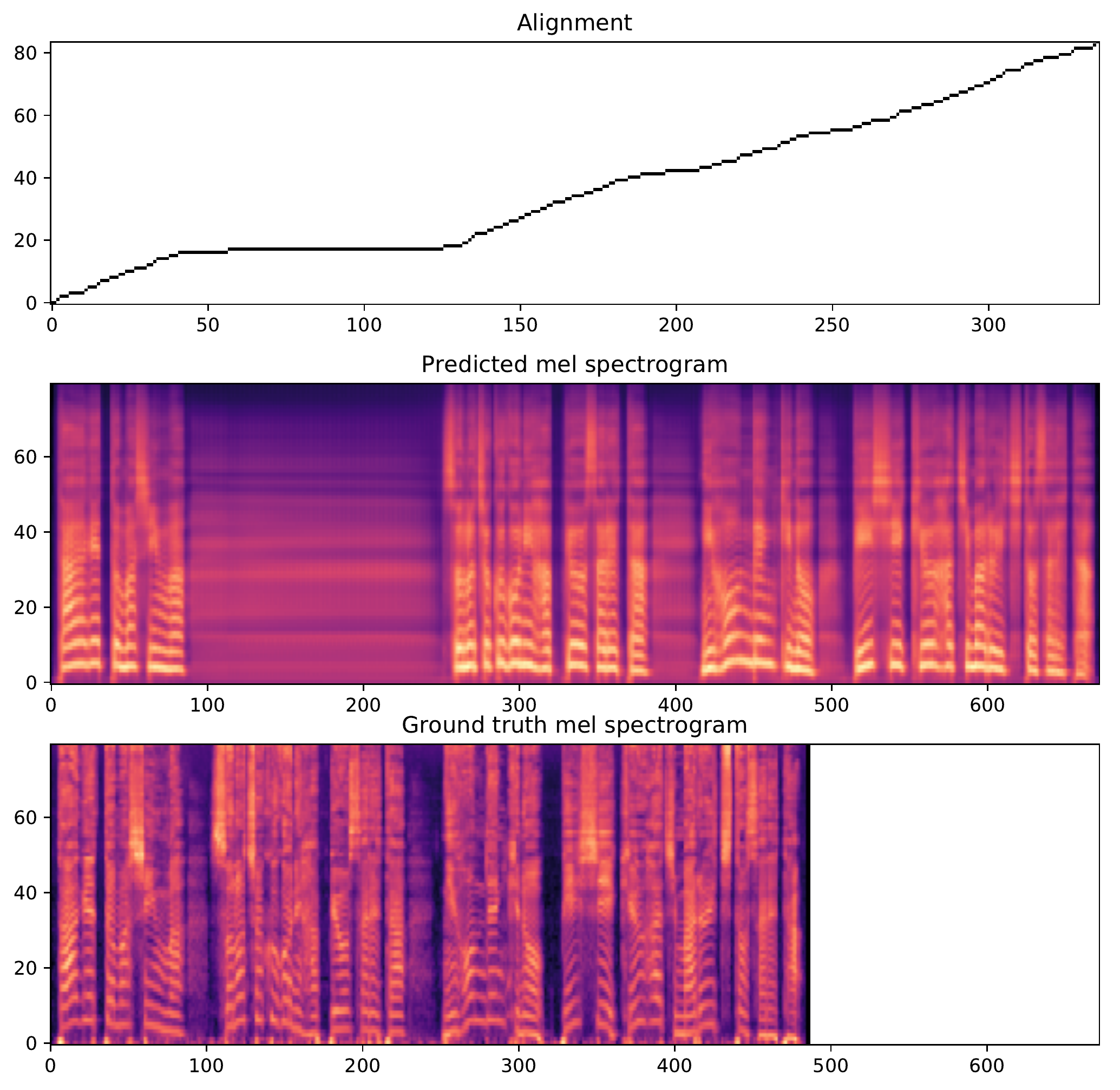}
  \caption{A sample that shows overestimation of pause duration.}
  \label{fig:alignment-error}
\end{figure}

To understand the reasons, we did a simple investigation of generated audio files and discovered out our model overestimated the phoneme duration, especially for pauses within a sentence.  Pauses had much longer duration than other phonemes, but, its acoustic features had less useful information; hence, deciding when \texttt{Shift} transition should be made would be difficult. Figure \ref{fig:alignment-error} shows an example sample that has alignment error of overestimation of pause duration. In fact, the MOS score of sentences that did not include pauses was $2.75 \pm 0.06$ and $3.24 \pm 0.05$ for SSNT and Tacotron, respectively, and the score difference was smaller. We also discovered that our method needed longer training time due to the marginalization process. The performance might be improved after using sufficiently long training periods. 

\subsection{Discussions}
\label{Effect of hard monotonic alignment}

%\textcolor{red}{(This section needs to be improved.)}

We designed the alignment structure to be hard and monotonic, which enabled us to avoid some alignment errors that are commonly observed in soft-attention-based approaches. Such alignment errors include muffling, skipping, and repeating. 
%, early and late termination. 
Muffling error is caused by an attention distribution without a sharp mode, skipping error is caused by discontinuous attention, and repeating error is caused by a repeat of backward jumps of the mode of attention. 
%The early and late termination errors are caused by wrong stop decision from stop flag prediction. 
%The early termination results in cutting off of tail part of transcription, and the late termination sometimes causes another error like the repeat error. 
These errors were not observed in our method because they could not happen by definition. 

However, we still observed different types of alignment errors in our samples, such as prolongations of vowel duration. We also observed that even though the alignment looked acceptable, that is, a monotonic increase occurred, wrong phonemes were sometimes generated due to a poorly trained model. When the alignment was not properly specified, its emission probability could be learned from different acoustic segments. This is a disadvantage of SSNT's hard alignments. However, we also found an advantage with the hard alignment. From our informal listening of generated speech, compared with soft-attention-based methods, speech generated from our method tended to have relatively distinct pronunciation. 

\section{Conclusions}

We proposed a new method for end-to-end TTS without soft attention. In contrast with soft attention based methods, our method has a simpler architecture and ensures monotonic alignment structure by design. Our method represents an alignment variable as a latent variable, and the model can be trained by maximizing the total probability that can be derived by marginalizing the latent alignments. During inference, the alignment variable can be randomly sampled from the learned distribution, and the inference simply stops when the alignment reaches the final input position.

Thanks to the design of hard and monotonic alignment, our proposed method could avoid some alignment errors that were commonly observed in soft attention based approaches. Our method also replaced many engineering features in soft attention based approaches with a probabilistic approach, which included advanced attention mechanisms to enforce a monotonic alignment, stop flag prediction network, and nondeterministic inference by dropout.

Although our analysis revealed that some generated speech contained another type of alignment errors, and although a subjective evaluation showed that the quality of synthetic speech from our system was not yet competitive with soft attention based methods on the reading speech corpus yet, our other research showed the effectiveness of the proposed method on a verbal performance corpus that is much more challenging data than reading speech \cite{KatoS2019rakugo}. 
%in that alignments predicted by soft-attention were quite unstable. 
%Further improvement of our proposed method is required for reading corpus. 

Our future work includes performance optimization for faster training of SSNT-based TTS. We expect fast training will help to reduce the remaining alignment errors induced by insufficient training time. In addition, we plan to use a more complex probability distribution function for the emission probability of the acoustic features. In this study, we chose the isotropic Gaussian distribution for the emission probability. This was not an ideal choice because the target mel spectrogram had a clear correlation across frequency dimensions. We expect that full covariance modeling will enable random sampling of acoustic features and that a sufficiently complex probability distribution will lead to higher quality generated speech.

\section{Acknowledgements}
This work was partially supported by a JST CREST Grant (JPMJCR18A6, VoicePersonae project), Japan, and by MEXT KAKENHI Grants (16H06302, 17H04687, 18H04120, 18H04112, and 18KT0051), Japan. The numerical calculations were carried out on the TSUBAME 3.0 supercomputer at the Tokyo Institute of Technology.

\bibliographystyle{IEEEtran}
\bibliography{BIB}

\end{document}